\documentclass[%
 reprint,
 amsmath,amssymb,
 aps,
prd,
]{revtex4-2}

\usepackage{graphicx}
\usepackage{dcolumn}
\usepackage{bm}
\usepackage{blindtext}
\allowdisplaybreaks[4]
\usepackage{dsfont}
\usepackage{units}
\usepackage{ulem}
\usepackage{physics}

\begin{document}


\title{Resolution of spin crisis, and notes on the Bjorken sum rule, anomaly and constituent quark.}

\author{J. Pasupathy}
\email{jphome@gmail.com}
\affiliation{Centre for High Energy Physics, Indian Institute of Science, Bengaluru-560012, India}
\author{Janardan P. Singh} 
\email{janardanprasad.singh@gmail.com}
\affiliation{Physics Department, Faculty of Science, The Maharaja Sayajirao University of Baroda, Vadodara-390002, Gujarat, India}

\begin{abstract}
 It is shown that the widely used parton model expression for $ g_1$ in polarized proton-lepton scattering is incorrect as it ignores gluon-quark spin entanglement. Therefore, there is no spin crisis. 
A brief summary of results of the theoretical evaluation of non-octet axial vector current
renormalization constants and anomaly -anomaly vacuum correlator is given. It suggests that anomaly plays an
important role in the transformation of current quarks to constituent quarks and chiral
symmetry breaking. 
\end{abstract}

\maketitle

\section{Introduction}
In this work, we first give in Section 2 a brief description of deep inelastic polarized proton-lepton scattering. We show that in the Bjorken limit, where a parton description is appropriate, the function $g_1 (x)$ cannot be written using quark helicity densities. The reason for this is explained using simple examples in quark and gluon helicity addition. Although the gluon does not interact with photon, the gluon helicity determines quark helicity because the total must add up to proton spin.  Taking into account the entanglement between gluon and quark helicity, we find the widely used parton expression for $g_1$ is invalid. What is more, $g_1(x)$  cannot be determined  using only parton ditribution probability functions.  Section 3 is on the theoretical determination of the non-octet axial vector current renormalization constant  and its comparison with  experiment. The importance of the axial anomaly is stressed. In theoretical calculations like QCD sum rules, taking $m_u =m_d=m_s =0$ (SU(3) symmetric theory)  is a good approximation. This is because the constituent masses, which are of order 300 MeV, are significantly larger than current quark masses. The latter can be set equal to zero first and then taken into account as a correction.  In Section 4 the issues involved in deriving the proton wave function as a superposition of bare quanta of quarks and gluons is discussed.  Understanding the constituent quark structure is a necessary first step. While valuable results have been obtained using Bogoliubov-Valatin transforms from the current quark into the constituent quark, they are inadequate. Phenomenology requires that we should find a wider class of transforms that includes transverse gluons.

\section{Polarized proton scattering}
Deep inelastic scattering by polarized lepton on polarized nucleon target has been intensely studied for many years now. For recent reviews see \cite{Bass, Gross, Riedl}. An earlier work by Kuhn, Chen and Leader\cite{Kuhn} has an excellent review of both theory and experiments. Following Ref.\cite{Kuhn} the difference of cross-sections with opposite target spins are given by 
\begin{equation}
\begin{aligned}
\Big[\frac{d^2\sigma}{d\Omega dE'}(k,s,P,-S;k')-\frac{d^2\sigma}{d\Omega dE'}(k,s,P,S;k')\Big] \\
=\frac{\alpha^2}{2Mq^4}\frac{E'}{E}4L^{(A)}_{\mu\nu}W^{\mu\nu(A)}
\end{aligned}
\end{equation}
Here E, E’ are the incident and scattered lepton energies in lab frame, and $\nu=E-E'$, $q_{\mu}$ is virtual photon four-momentum .  The leptonic tensor is
\begin{equation} L^{(A)}_{\mu\nu}(k,s;k')=m\epsilon_{\mu\nu\alpha\beta}s^{\alpha}(k-k')^{\beta} \end{equation}
$s^{\alpha}$ is lepton spin polarization vector. The hadronic tensor written in term of invariant $ G_1$ and $G_2$ is\\
\begin{equation}
\begin{aligned}
\frac{1}{2M}W^{(A)}_{\mu\nu}(q;P,S)=\epsilon_{\mu\nu\alpha\beta}q^{\alpha}\Big\{MS^{\beta}G_1(P.q,q^2) \\
+[(P.q)S^{\beta}-(S.q)P^{\beta}]\frac{G_2(P.q,q^2)}{M}\Big\}
\end{aligned}
\end{equation}
$S_{\mu}$ is proton spin polarization vector, M is proton mass, $P_{\mu}$ its momentum. In the Bjorken limit  with x fixed,
the functions $g_1$ and $g_2$ are given by
\begin{equation}
\begin{aligned}
\lim_{Bj}\frac{(P.q)^2}{\nu}G_1(P.q,Q^2)=g_1(x)\\
\lim_{Bj}\nu(P.q) G_2(P.q,q^2)=g_2(x)
\end{aligned}
\end{equation}
The functions $g_1$ and $g_2$ scale approximately with very slow logarithmic variation.
The Bjorken sum rule \cite{Bjorken}  relates the integral of $g_1(x)$ to axial vector matrix elements
\begin{equation}
\begin{aligned}
\hat{\Gamma}^p_1\equiv  \int_{0}^{1}g_1(x) dx =\frac{1}{2}(\frac{4}{9}G_U+\frac{1}{9}G_D+\frac{1}{9}G_S) \\
= G_{Bj}
\end{aligned}
\end{equation}
where\\
\begin{equation}\langle P|\bar{u}\gamma_{\mu}\gamma_5u|P\rangle=G_U\bar{v}\gamma_{\mu}\gamma_5v \end{equation}
 The proton spinor $v$ is normalized as
\begin{equation}\bar{v}v=2M \end{equation}
$G_U$ is the renormalization constant for u quark axial vector current, and for d and s quarks similar definitions hold. The axial current renormalization constants for octet currents matrix elements used in semi-leptonic decays of  baryons have been known for a long time. The constant $G_{Bj}$ cannot be expressed as a linear combination of the octet currents renormalizations constants only. One needs to know the singlet current renormalization constant, which is not available from semi-leptonic decay data. The first experiment on polarized proton scattering by \cite{Ashman}  found the value
\begin{equation}\hat{\Gamma}^p_1= 0.114 \pm 0.012 \pm 0.026  \end{equation}
Following this result, Leader and Anselmino \cite{Leader88}  reached the conclusion that very little of the proton's spin is carried by quarks. The analysis is easily explained in simple terms in Ref. [4] which we follow.  According to  the parton model expression \cite{Feynman, Altarelli}
\begin{equation}
\begin{aligned}
g_1(x)=\frac{1}{2}\sum_{j} e^2_j\big[\Delta q_j(x)+\Delta \bar{q}_j(x)\big] \\
\Delta q(x)=q_+(x)-q_-(x) \\
                      q(x)=q_+(x)+q_-(x) 
\end{aligned}
\end{equation}
Here $q_{\pm}$ are the number densities of quarks whose spin orientation is parallel or antiparallel to the longitudinal  spin direction of the proton and $q(x)$ is the usual parton density. Using SU(3) decomposition one can write
\begin{equation} g_1(x)=\frac{1}{9}\big[\frac{3}{4}\Delta q_3(x)+\frac{1}{4}\Delta q_8(x)+\Delta \Sigma(x)\big] \end{equation}
where
\begin{align}
\Delta q_3(x) &= (\Delta u(x) + \Delta \bar{u}(x)) \nonumber \\
              &\quad - (\Delta d(x) + \Delta \bar{d}(x)) \nonumber \\
\Delta q_8(x) &= (\Delta u(x) + \Delta \bar{u}(x)) \nonumber \\
              &\quad + (\Delta d(x) + \Delta \bar{d}(x)) \nonumber \\
              &\quad - 2(\Delta s(x) + \Delta \bar{s}(x)) \nonumber \\
\Delta \Sigma(x) &= (\Delta u(x) + \Delta \bar{u}(x)) \nonumber \\
                 &\quad + (\Delta d(x) + \Delta \bar{d}(x)) \nonumber \\
                 &\quad + (\Delta s(x) + \Delta \bar{s}(x))
\end{align}
Take the first moment of $g_1(x)$ :
\begin{equation}
\hat{\Gamma}^p_1=\frac{1}{9}\big[\frac{3}{4}a_3+\frac{1}{4}a_8+a_0\big]
\end{equation}
where
\begin{equation}
\begin{aligned}
a_3=\int_{0}^{1} dx \Delta q_3(x)\\
a_8=\int_{0}^{1} dx \Delta q_8(x)\\
a_0=\int_{0}^{1} dx \Delta \Sigma(x)
\end{aligned}
\end{equation}

Now using the values $a_3=G_U-G_D$, $a_8=G_U+G_D-2G_S$ and the experimental value of $\hat{\Gamma}^p_1$, they find $a_0 \cong 0$ while according to Leader and Anselmino \cite{Leader88} one should expect a value close to $1/2$, the proton spin.
 This is the spin crisis.
 
It is not difficult to see the shortcoming of Eq. (9). It tacitly assumes that when the target proton is polarized say in positive z direction with $S_{z} = 1/2$, the sum $s_z$  of quarks and antiquarks spin component in z-direction add up to $+1/2$, the proton spin. This need not be the case, since this ignores gluon spin. In reality, the quarks which have spin $1/2$ and gluons with spin $1$ are entangled. It is the vector sum of quark plus gluon spins that must add up to $1/2$, the proton spin. That is, although the gluons do not interact with the photon, the gluon spin sum dictates the value of the sum of quark spin. 

To appreciate this point which is quite simple, consider an example of 3 massless fermion and 1 massless boson all moving along the z-axis. The fermions have helicity either plus or minus half. For a scalar boson helicity is zero while for the vector massless boson the helicity is either plus or minus one. Let us consider the number of ways in which the total helicity adds up to plus half, first for scalar case. It is evident that there is only one way, apart from permutations between the two fermions that have helicity $+\frac{1}{2}$, with the third fermion helcity fixed at $-\frac{1}{2}$. The scalar boson is a mute spectator in this case. On the other hand, the vector case is quite different. Let the total helicity be plus half as before. The boson helicity has two possibilities +1 and -1. In the case where the boson helicity is +1, two fermions have helicity $-\frac{1}{2}$ while the third has $+\frac{1}{2}$ - the exact opposite of the scalar case. If the boson helicity is -1, then all the three fermions have helicity $+\frac{1}{2}$. This  fermion configuration does not even exist in the scalar case.

Now let us recall the main points of Feynman’s parton model \cite{Feynman}. 
\begin{itemize}
    \item There is an underlying field theory which describes the proton.
    \item In such a field theory, the proton has a wave function which can be expressed as a superposition of Fock states with varying number of quanta of the theory.
    \item Scaling in deep inelastic  scattering of lepton on proton  is due to scattering of bare quark with virtual photon. 
\end{itemize}

Now in QCD, gluons are massless and light quark masses can be ignored when studying
deep inelastic scattering of the proton. We can, by following Feynman’s ideas \cite{Feynman1}, write  the proton state as a superposition of Fock states of perturbation theory.
                                                                 
\begin{eqnarray}
|proton\rangle=\sum_{m,n} C_{m,n,p^r,s_r}|q,\bar{q},g \rangle _{m,n}\\
|q,\bar{q},g \rangle _{m,n} =\Pi a^{\dagger}(p^i,s_i) \Pi d^{\dagger}(p^j,s_j) \Pi b^{\dagger}(p^k, s_k)|0 \rangle
\end{eqnarray}
 Here,  $|0 \rangle$ is the perturbation theory vacuum, $a\dagger$, $d\dagger$ and $b\dagger$ are the creation operators for the gluon, antiquark and quark respectively. We have suppressed favour and colour  indices.   Eq.(15)   denotes a state with m+3 quarks $q$, m antiquarks $\bar{q}$, n gluons, the index k for quarks runs from 1 to m+3, the index j for antiquarks runs from m+4 to 2m+3, the index i for gluon runs from 
1 to n.  There are n+2m+3 partons with different momenta and spin component $(p^r,s_r)$ .   
The coefficient $C_{m,n,i,j,k,p^r s_r}$  depends on   quark-antiquark's spin component (plus or minus half), flavor, color, momenta and gluons helicities (plus or minus one), color, momenta.  Each term in this expansion of  the wave function must be color singlet, the sum of parton momenta equals the proton momentum, sum of  gluon and $q (\bar{q})$ spins add up to proton spin. Feynman viewed the collision with the photon in a frame where proton has momentum tending to infinity in the z direction. Scaling is due to virtual photon hitting a bare quark or antiquark with momentum $xp_z$.  The probability of the event is proportional $|C_{m,n,i,j,k,p^r s_r}|^2$.  The coefficients  $C_{m,n,i,j,k,p^r s_r}$  can not be calculated in perturbation theory [Cf. sec. 4].

Now consider polarized proton lepton scattering. The experiment measures the difference between the cross section when proton spin S is parallel to lepton spin s and when S and s are  anti parallel. Let the spin of the proton $S_z =1/2$. Consider the wave function component with a specific value for n and m. If n is odd clearly the total gluon helicities cannot add up to zero. Among various combinations of gluon and quark spins, take the case where n-1 gluons helicities add to zero and a lone gluon has helicity +1. Then the total $q + \bar{q}$ helicities must add up to minus half. That is, even though the gluon does not interact with photons, gluon helicity +1 compels the $q + \bar{q}$ helicities to add up to -1/2 that is antiparallel to proton spin. On the other hand, if the lone gluon has helicity $- 1$ then quark plus antiquark helicities should add to 3/2, that is, an excess of three $q(\bar{q})$ helicities parallel to proton with the remainder adding up to zero.

Contrast this with a scalar gluon theory. Then only quarks have spin. The 3+2m quarks and antiquarks spins must add up to ½.  When proton spin $S_z =1/2$, there is an excess of one quark or antiquark which is spin  parallel to proton while the others add up to zero. The scalar gluon number n is irrelevant. We can sum over n trivially and average over m, taking into account the u-quark  charge of $2/3$ and d-  and s- quark charge $-1/3$ to arrive at the expression for $g_1$ with  $\Delta q(x)$ and $q(x)$ as given in Eq. (9).

In QCD  the situation is drastically different. We already saw above that even a special case when n is odd, gluon helicity sum is plus +1,  quark helicity combination is -1/2  and when  gluon helicity  is  -1  quark helicities add up  3/2. Further even with a specific value for index m and n fixed there are  large number of ways in which the gluons and the quarks can partition their helictites to meet sum plus or minus half. Varying m and n introduces more complex combinations. The polarization asymmetry in proton lepton scattering in the partonic context, depends on quark helicities parallel and antiparallel to lepton helicity. There is no regular pattern either within m and n as we vary partitions or as we vary m and n. Therefore it  is impossible to write for $g_1(x)$  any expression using parton  distribution functions in QCD. Thus we have an ironic situation - Eq.9 is valid in scalar theory which has no asymptotic freedom 
while in QCD in which gluon spin responsible as asymptotic freedom, Eq.9 is invalid precisely
because the gluon has spin. Therefore the analysis given in \cite{Leader88, Kuhn} using \cite{Feynman, Altarelli} is incorrect. 

It is useful to remark on the difference between energy momentum and Bjorken sum rule.  A) The former involves unpolarized cross-sections while the latter involves the polarized ones, which means the difference of parallel and anti-parallel cross-sections B) Eigenvalues of energy and momentum have continuous spectra while spin values are discrete, half or one. So, to summarize:\\
1) There is no spin crisis. \\
2)  While the parton idea is valid, one should not expect every experimentally   measured quantity to be  expressible by the parton distribution function. \\
3) The wave function, as given by Eqs. (14) and (15), is fundamental to understand the proton.  \\ 
We will consider phenomenology elsewhere. In the next section we discuss the theorertical determination of renormalization constants, both for octet and non-octet currents.

\section{Anomaly in the context of Bjorken sum rule}
Consider the QCD Lagrangian
\begin{equation}\mathcal {L}_{QCD}=\sum_{j=u,d,s}\bar{q}^j(i\not{D}-m_j)q^j -\frac{1}{4}G_{\mu\nu}^aG^{a\mu\nu}  \end{equation}
The covariant derivative is $D_{\mu}=\partial_{\mu}-ig_s A^a_{\mu}T^a$, $g_s$ is the gauge coupling constant  and the field strength tensor is $G^a_{\mu \nu}=\partial_{\mu}A^a_{\nu}-\partial_{\nu}A^a_{\mu}+  f^{abc}g_sA^b_{\mu} A^c_{\nu}$. $m_j$ are the quark masses \cite{Zyla}:
\begin{align}
m_u &= (2.16 \pm 0.38)\ \text{MeV}, \nonumber \\
m_d &= (4.67 \pm 0.32)\ \text{MeV}, \nonumber \\
m_s &= (93.0 \pm 0.8)\ \text{MeV}
\end{align}
The divergence of axial vector current has an anomaly \cite{Bell, Adler, AdlerB} :
\begin{equation} \partial^{\mu}\bar{\Psi}_q\gamma_{\mu}\gamma_5 \Psi_q=2im_q\bar{\Psi}_q\gamma_5 \Psi_q +\frac{\alpha_s}{4\pi}G^a_{\mu\nu}\tilde{G}^{a\mu\nu} \end{equation}
where  and  $\alpha_s=\frac{g_s^2}{4\pi}$
 
It was pointed out in \cite{Gross77}  that if the anomaly is neglected then, one expects a large violation of isospin symmetry of the order of $(m_d- m_u)/ (m_d +m_u)$ being  around 30\% in Bjorken  sum rule.  Since the first experiment by Ashman et.al. \cite{Ashman} many experiments have been done including those from which neutron data is also available. From \cite{Adolph} we learn 
\begin{equation}\hat{\Gamma}^p_1  - \hat{\Gamma}^n_1  = 0.181 \pm 0.008 \pm 0.014 \end{equation}
which agrees with the Bjorken’s prediction and there is no sign of large violation of isospin in data.  Now consider the  Ellis-Jaffe sum rule \cite {Ellis} which uses the value of $G_U +G_D -2G_S$,  the renormalization constant of octet current whose divergence is anomaly free, from the knowledge of semi-leptonic decays, and  also assumes $G_S =0$ which is  apparently justified by OZI rule. But if the strange quark current is dropped then $\bar{u}\gamma_{\mu} \gamma_5 u+\bar{d}\gamma_{\mu} \gamma_5d$ has  an anomaly. If anomaly is neglected, as we noted above, one expects large violation of isospin and flavor SU(3) symmetry. Thus, the Ellis-Jaffe sum rule \cite {Ellis}, which uses two incompatible assumptions is bound to fail as confirmed by the experiment \cite{Adolph}. Inclusion of anomaly is important to understand the experimental data.

We now give a quick summary of results for the axial renormalization constants using QCD Sum Rules which were calculated four decades ago.  The study of baryon properties using QCD Sum Rules was pioneered by Ioffe who introduced the proton current operator\cite{Ioffe81}
\begin{eqnarray}
\eta=\epsilon^{abc}[u^a(x)C\gamma_{\mu}u^b(x)]\gamma^{\mu}\gamma_5d^c(x)\\
\Pi=i\int d^4x e^{ipx}\langle \phi|T\{\eta(x),\bar{\eta}(0)\}|\phi\rangle
\end{eqnarray}
where C is the charge conjugation and $|\phi \rangle$ is the physical vacuum. It has the property \cite{Gell-Mann}
\begin{equation} \langle \phi| \bar{q}q|\phi \rangle=\kappa \simeq-(240 MeV)^3 \end{equation}
This quantity is non-zero signaling break down of chiral symmetry of QCD. Besides $\kappa$ there are  other Lorentz invariant fields operators, with non- zero vacuum expectation values like $\langle\phi|\alpha_s G^a_{\mu\nu}G^{a\mu\nu}|\phi\rangle$.  Using these values and operator product expansion, Ioffe \cite{Ioffe81}   computed the nucleon mass. Skipping the details, the important point is that most of nucleon mass is given by chiral symmetry breaking and quark masses can be neglected, 
\begin{equation} m_N\thicksim (-8\pi^2 \kappa)^{1/3}\simeq 1 GeV     \end{equation}
Belyavev and Ioffe \cite{Belyaev83}  extended the nucleon calculations to baryon octet masses and spin-3/2 beryon-resonance masses by including the quarks masses with good results. Therefore, taking the light quark masses as zero is a good starting point for calculations. Apart from mass, other physical quantities like magnetic moments, axial vector current renormalization constants for the octet baryon have been calculated using QCD sum rules. This is done by adding an interaction term to $\mathcal {L}_{QCD}$. 
\begin{equation} \mathcal {L}_{int}=J_{\mu}A^{\mu}    \end{equation}       
where $A_{\mu}$   is   a constant external magnetic field and  $J_{\mu}$  is electromagnetic current for calculating magnetic moments, while  $A_{\mu}$ is a constant external pseudovector field and  $J_{\mu}$  is the axial vector current for computing $G_A$ etc. 
In the presence of the external field $A_{\mu}$, the vacuum responds similar to materials in condensed matter theory. That is, vacuum has non-Lorentz  invariant  expectation values proportional to external field. Such calculations for magnetic moments were first done by Ioffe and Smilga \cite{Ioffe84}, and Balitsky and Yung \cite{Balitsky}.  Calculations   for axial vector current renormalization constants were done by many other authors \cite{ Belyaev1, Belyaev85, Chiu, Gupta,  Pasupathy}. In the context of the Bjorken sum rule \cite{Bjorken} one needs to determine $G_0=G_U+G_D$. Gupta, Murthy and Pasupathy \cite{Gupta}  using the earlier calculations of  Chiu, Pasupathy and Wilson \cite{Chiu} obtained a value 
\begin{equation} G_0  =  G_U + G_D  \simeq 0.35   \end{equation}
while in \cite{Pasupathy}  the authors   obtained for $G_0$  the lower value of  0.22.  It was noted, results for $G_0$ were sensitive to the external induced vacuum susceptibility. On the other hand, Singh, using a different approach \cite{singh2015} found $G_0\approx0.46$. Now that $G_0$ has been determined by experiment \cite{Adolph} (see Table 3) we can look at the data from a different perspective. We have
\begin{eqnarray}
\langle P|\bar{u}\gamma_{\mu}\gamma_5 u-\bar{d}\gamma_{\mu}\gamma_5 d |P\rangle=G_A\bar{v}\gamma_{\mu}\gamma_5 v \\
\langle P|\bar{u}\gamma_{\mu}\gamma_5 u+\bar{d}\gamma_{\mu}\gamma_5 d |P\rangle=G_0\bar{v}\gamma_{\mu}\gamma_5 v
\end{eqnarray}
Using Eq. (18) we get
\begin{align}
\langle P|\,&(m_u + m_d)(\bar{u}i\gamma_5 u - \bar{d}i\gamma_5 d) \nonumber \\
&+ (m_u - m_d)(\bar{u}i\gamma_5 u + \bar{d}i\gamma_5 d)\,|P\rangle \nonumber \\
&= G_A \cdot 2M\, \bar{v}i\gamma_5 v
\end{align}

\begin{align}
\langle P|\,&(m_u - m_d)(\bar{u}i\gamma_5 u - \bar{d}i\gamma_5 d) \nonumber \\
&+ (m_u + m_d)(\bar{u}i\gamma_5 u + \bar{d}i\gamma_5 d) + 4Q\,|P\rangle \nonumber \\
&= G_0 \cdot 2M\, \bar{v}i\gamma_5 v
\end{align}

where $Q(x)$=$\alpha_s G^a_{\mu\nu}(x)\tilde{G}^{a\mu\nu}(x)/(8\pi)$. There are three matrix elements  $\langle P|\bar{u}i\gamma_5 u-\bar{d}i\gamma_5 d|P\rangle$, $\langle P|\bar{u}i\gamma_5 u+\bar{d}i\gamma_5 d|P\rangle$  and $\langle P|Q(0)|P\rangle$  to be computed. This can be done by the external field method  briefly mentioned earlier \cite{JPS}. 

The correlator $\int d^4x e^{ipx} \langle \phi|T(Q(x),Q(0))|\phi \rangle$  has been widely studied in various contexts. 
In \cite{Ioffe98, Narison} it is discussed in the context of the so-called missing spin of the proton. 
In \cite{JP} this correlator is calculated in two different ways, one using experimental masses of light quarks and mixing angles of pseudoscalars and another theoretical one where all the quark masses are zero. In the theoretical case where the quark masses are zero, the octet of Nambu-Goldstone bosons are massless while the singlet has mass. By relating the two cases Ref. \cite{JP} obtained       
 \begin{equation}  M_{\eta_0}=723 MeV \end{equation}
 and corresponding 
\begin{equation} F_{\eta_0}=178 MeV \end{equation}
 ($f_{\pi}=133 MeV$). We note that the singlet mass is close to twice the constituent quark mass. From Eqs. (30) and (31) we have
\begin{equation} (M_{\eta_0} F_{\eta_0})^2/12=(193 MeV)^4 \end{equation}
Now according to Witten-Veneziano (WV) relation \cite{Witten, Veneziano} the left hand side equals the topological  susceptibility  in pure  gauge SU(3) theory (no quarks) )
\begin{equation} (M_{\eta_0} F_{\eta_0})^2/12=\langle \text{gauge vac}|\int d^4x T(Q(x)Q(0))| \text{gauge vac} \rangle \end{equation}
The right hand side was calculated using lattice gauge theory \cite{Luigi, Alles}. Authors of Ref. \cite{Luigi} obtained the value $(191\pm 5 MeV)^4 $, while \cite{Alles} obtained $(173\pm 0.5 \pm 1.2^{ +1.1}_{-0.2} MeV)^4 $ in good agreement with Eqs. (30-32). Assuming the validity of WV relation for  single flavor ($n_f$=1) and two flavors ($n_f$=2), further assuming that flavor singlet $F_{\eta_0}$ is independent of number of flavors as well we have : 
\begin{equation}  M_{\eta_0}=417 MeV (n_f=1),  M_{\eta_0}=590 MeV (n_f=2) \end{equation}

\section {Proton parton wave function}
We return to a discussion of the proton parton wavefunction Eqs. (14) and (15).  While we have no road map to derive the wavefunction from first principles, we hope the following remarks are useful.  The quark model predates QCD. Many useful results on baryon propertiess were obtained using non-relativistic wavefunctions of constituent quarks U,D,S. In the early  years of the Parton Model  Altarelli, Cabibbo , Maini and Petronzio \cite{Altarelli69} constructed a parton distribution function as follows. 1) From the quark model wave function calculate the momentum distribution functions of U and D quarks. 2)  Go to the frame where $P_z$ tends to infinity and express the momentum distributions in term of $x_i =p_{z_i}/P$ and transverse momenta. 3) Since the U, D quarks are themselves composite, resolve them  into a distribution of bare $u, d, s$ quarks, antiquarks plus gluons  using phenomenological considerations like Regge theory. 4) Folding this distribution with U, D distributions in step two they obtained proton’s  parton distribution function.  Although  no attempt was made  to derive the constituent quark structure  from basic QCD theory, two useful facts emerged from their work. First, one should include gluons to agree with experiment, second  include $s\bar{s}$ alongwith $u\bar{u}$ and $d\bar{d}$ in the transform from current quarks  to constituent U and D quarks.

In QCD, it is known that the bare vacuum is unstable. From phenomenology, chiral symmetry is spontaneously broken.   Further, if the gluon-quark coupling is sufficiently strong, quark- antiquark pairs can condense to lower ground state energy.  In the works of Refs. \cite{Finger82, AmerL, Govaerts, AdlerD}  current quark to constituent quark transformation was made imitating BCS theory of superconductivity \cite{Schrieffer}. Following Adler and Davis \cite{AdlerD}, instead of the complete Hamiltonian one can start with an effective Hamiltonian for color quark charge densities only
\begin{equation} H_{eff}=\bar{q}\gamma .(-i\nabla )q-2\pi \sum_{a}\bar{q}\gamma_0\frac{1}{2}\lambda^aq\frac{\alpha_s}{\nabla^2}\bar{q}\gamma_0\frac{1}{2}\lambda^aq \end{equation}
The summation is over the color index \textit{a}, and $\alpha_s=g_s^2/(4\pi)$.  To  find the minimum of $H_{eff}$, the trial wave function for the ground state is taken to be a  coherent superposition of $q\bar{q}$ pairs :

\begin{align}
|\Psi \rangle = \frac{1}{N(\Psi)} \prod_{p,s,a} \big[\,1 
&- s\,\Psi(p)\,\tau\, b^{(a)\dagger}(\mathbf{p},s) \nonumber \\
&\times d^{(a)\dagger}(-\mathbf{p},s)\,\big]\,|0\rangle, \quad p = |\mathbf{p}|
\end{align}

with $b^{(a)\dagger}(\mathbf{p},s)$ and $d^{(a)\dagger}(\mathbf{p},s)$ being the creation operators for a quark and antiquark with momentum $\mathbf{p}$, color index \textit{a} and helicity s. In Eq. (36) $\tau $ is the volume of an elementary cell in momentum space and $ \Psi(p)=  \Psi(p)^*$ is the momentum space gap function. Skipping details of renormalization and numerical solution of the gap equation, we learn from Adler and Davis\cite{AdlerD} that  the gap function $ \Psi(p)$ exists even for a confining potential, and further $ \Psi(p)$ rapidly tends to zero for large p. Further, a knowledge of  $ \Psi(p)$  provides a resolution of the constituent quark state into  a superposition of current quark-anti quarks states. The chiral symmetry  breaking expectation value found by Adler and Davis is
\begin{equation} \langle \Psi |\bar{q}q|\Psi\rangle \cong(-95 MeV)^3,  \end{equation}
They also obtained for the constituent quark mass a value $\cong 70 MeV$ from the shape of the gap function   near p$\approx 0$. These are quite low compared to the phenomenological value as discussed in section 3, Eq.(22).  One of the reasons, as   listed by them, could be the neglect of non-planar diagrams, that is, the anomaly in their derivation of renormalized gap function $ \Psi(p)$. Another reason is that eqn. (37) is calculated for a single quark in \cite{AdlerD} whereas the phenomenological value in eqn (22) is for the three u, d, s quarks. Using WV formula \cite{Witten, Veneziano} and assuming its validity for single flavor we expect the pseudoscalar mass 
\begin{equation} M_{\eta_0}\cong 723/\surd{3} MeV \cong 417 MeV \end{equation}
This suggests that including three flavors, chiral symmetry breaking vacuum expectation value and the constituent mass value   can improve the calculation substantially.  Further, by including all the three quarks one can also understand how the constituent U quark state includes not only $\bar{u}u$ pairs, but also $\bar{d}d$ and $\bar{s}s$ pairs. But failure to include the gluonic partons, which is necessary for agreement with experiment, is a serious drawback. Construction of the QCD vacuum without quarks first and then including the quarks and chiral transformation later, appears necessary towards the goal of unraveling proton wavefunction Eqns. (14) and (15).

\bibliographystyle{unsrt}
\bibliography{references}

\end{document}